\documentclass[runningheads]{svmult}

\usepackage{makeidx}   
\usepackage{graphicx}  
\usepackage{subeqnar}  
\usepackage{multicol}  
\usepackage{physprbb}  
\makeindex             

\begin{document}
\title*{High redshift ellipticals: prospects for the VLT}
\toctitle{Focusing of a Parallel Beam to Form a Point
\protect\newline in the Particle Deflection Plane}
%
%
\titlerunning{High-$z$ ellipticals}
%
\author{Andrea Cimatti\inst{1}}
\authorrunning{Andrea Cimatti}
%
%
\institute{Osservatorio Astrofisico di Arcetri, Largo E. Fermi 5, I-50125,
Firenze, Italy}

\maketitle              

\begin{abstract}
The results and the present limits of the observations of high-$z$
ellipticals are discussed in the framework of VLT imminent and
future instruments.
\end{abstract}

\section{When and how did massive ellipticals form ?}
The formation of the present-day massive elliptical galaxies remains 
one of the most controversial issues of galaxy evolution and structure
formation. In CDM hierarchical merging models (e.g. \cite{k96,bau,bau2}),
 massive ellipticals form at relatively low redshift (e.g. $z<1$) through the 
merging of spiral galaxies. In such scenarios, massive, old and 
passively evolving ellipticals are extremely rare objects at $z>1$. 
In marked contrast, other scenarios suggested that massive ellipticals 
formed at higher redshifts (e.g. $z>2-3$) through an intense initial 
starburst event followed by pure luminosity evolution (PLE) of the 
stellar population to nowadays (e.g. \cite{egg,larson,arimoto}), thus
implying a constant comoving number density of passively evolving 
ellipticals at $z\sim 0$ and $z>1$. In this scenario, a substantial
number of extremely red objects (EROs) with the colors of  
an old stellar population at $z>1$ (e.g. $R-K>5.3$) and $r^{1/4}$
surface brightness profiles typical of dynamically relaxed
spheroidals is expected to be found in near-IR selected (i.e.
stellar mass selected) galaxy samples.

Imaging surveys with typical fields of $\sim$1-50 arcmin$^2$ provided very 
discrepant results. Some found a clear deficit of old ellipticals at 
$z\sim 1$ or $z>1$ compared to passive evolution models (e.g.
\cite{zepf,fra,bar,men,treu,rod}), whereas others found evidence 
for a constant comoving density (\cite{tot,ben,broad,schade,scod,im}).
A large part of the above discrepancies is certainly due to
the strong ERO angular clustering (i.e. field-to-field density
variations) that was discovered thanks to wider field surveys 
(\cite{daddi1,mc}). The results of such 
surveys suggest that the observed angular clustering is the 
signal of the underlying 3D large scale structure of 
massive ellipticals\cite{daddi3}, and showed that the surface 
density of $z>1$ passive elliptical {\it candidates} is consistent with 
that expected in PLE models, thus suggesting that most field 
ellipticals were fully assembled at least by $z\sim$2.5 (\cite{daddi2}).
However, follow-up observations are needed to confirm that most 
EROs are passive ellipticals because it is known that some EROs 
are dust-reddened starburst galaxies, thus representing a ``contamination'' 
in color--selected samples of elliptical candidates at $1<z<2$
(\cite{c98,c99,d99}).

\section{The results of VLT 1st generation instruments}

Fig.1 shows the spectrum of a $z\sim1.1$ passive elliptical
as observed with the ESO VLT equipped with the optical 
imager--spectrograph FORS2 (\cite{cim_k20}). No emission lines are present,
and the main features are the 4000~\AA~ continuum break together 
with strong CaII H\&K absorptions and other weaker absorption lines.
When such a spectrum is redshifted to $z>1.4$, the 4000~\AA~ and the 
CaII lines exit from the accessible optical spectral range and the 
redshift identification relies only on weak absorptions. Fig. 2 
shows an example of an elliptical candidate at $z\sim1.6$ that
can be taken as a clear example of the difficulties in 
identifying the nature and the redshifts of high-$z$ passive 
ellipticals (see also \cite{liu,sto}).

\begin{figure}
\begin{center}
\includegraphics[width=0.8\textwidth]{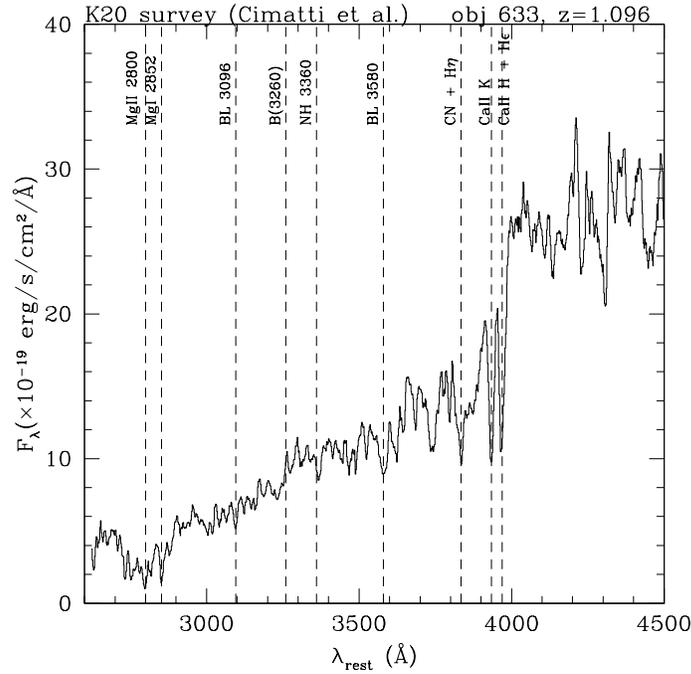}
\end{center}
\caption[]{ESO VLT+FORS2 spectrum of a $z\sim1.1$ elliptical with $R\sim23$
(from \cite{cim_k20})}
\label{eps1}
\end{figure}

When elliptical candidates are too faint for optical spectroscopy
and/or expected to be at $z>1.5$, the only possibility is to move
to near-IR spectroscopy in order to search for the 4000~\AA~ break
and the CaII H\&K lines redshifted at $\lambda_{obs}>1\mu$m.
However, the general faintness of the target continua makes
such observations difficult and time--consuming, and the
results of both Keck and VLT seeing--limited near-IR spectroscopy 
provided so far rather ambiguous results (\cite{soi,c99,glas}).

Despite the difficulties of optical and near-IR spectroscopy, 
a substantial number of $z>1$ passively evolving ellipticals
has been identified in recent surveys
\cite{spi,liu,cohen,cim_k20,dunlop,sto}. The inferred
ages of the stellar populations ($\sim 2-5$ Gyr) are consistent
with such galaxies being formed at remote cosmological epochs.
In addition, morphological studies based on HST imaging
further confirmed the existence of a population of dynamically
relaxed high-$z$ spheroids through the analysis of their surface 
brightness profiles (e.g. \cite{morio,treu,sti,rod}.

\begin{figure}
\begin{center}
\includegraphics[width=0.8\textwidth]{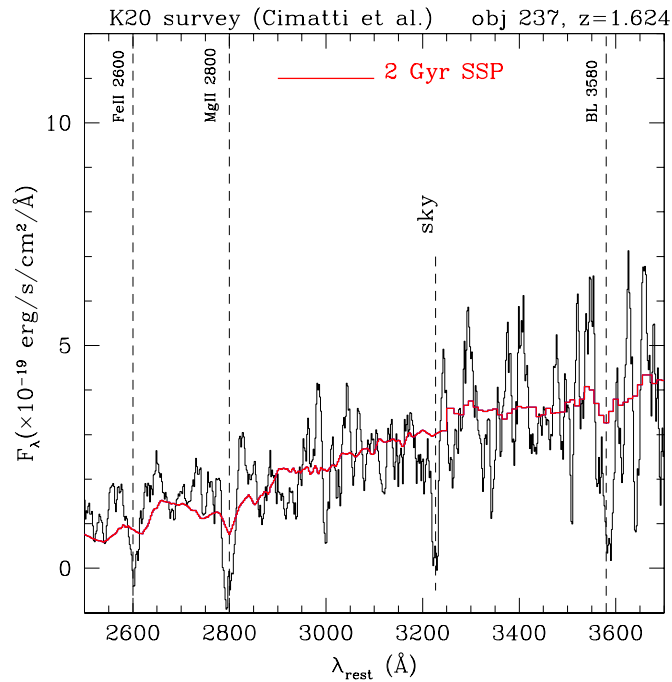}
\end{center}
\caption[]{ESO VLT+FORS2 spectrum of a $z\sim1.6$ elliptical candidate
with $R\sim25.9$ and $R-K=6.8$ plotted together with a Bruzual \& Charlot
2 Gyr old SSP model (from \cite{cim_k20}).}
\label{eps1}
\end{figure}

\section{Future prospects for the VLT}

Despite the successful observations of VLT, Keck and HST to 
uncover a substantial population of passively evolving ellipticals
at $z>1$, many questions need still to be answered before
drawing final conclusions on the formation and evolution
of massive galaxies: what is the fraction of ellipticals
as a function of $z$ and $K$ ? what are their masses, ages and 
redshifts of formation ? what is the evolution of their luminosity 
function and clustering with respect to model predictions ?  
Addressing the above questions requires the use of different
observing techniques, and three ideal VLT instruments could play
a crucial role:

\begin{itemize}

\item {\it AO--aided near-IR imager.} Such an instrument,
especially if aided by a laser guide star (LGS) system allowing
a flexible pointing of the telescope (i.e. not limited to targets
close to bright natural stars), would allow to study
the morphology of elliptical candidates not only to confirm their
nature by analysing their surface brightness profiles, but also
to derive radial color gradients and to study the evolution of
the fundamental plane (e.g. via the Kormendy relation) once the 
redshifts are known.

\begin{figure}
\begin{center}
\includegraphics[width=1.0\textwidth]{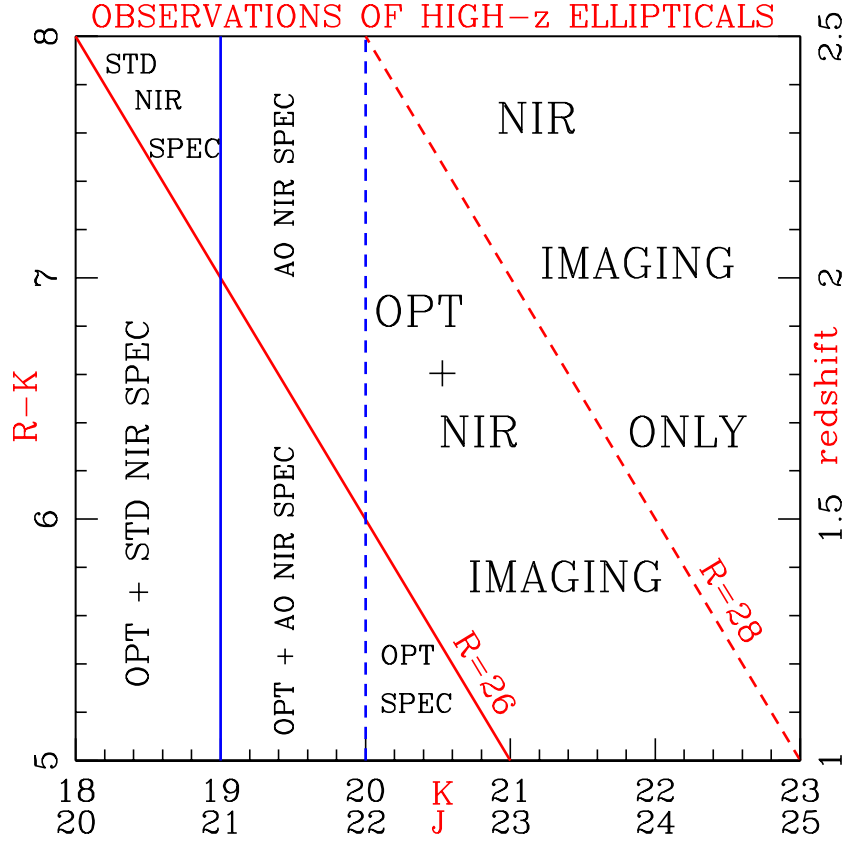}
\end{center}
\caption[]{The figure displays the different observing regimes for
high-$z$ passive ellipticals as a function of their colors, magnitudes
and redshifts. The X-axis shows the $J$ and $K$ magnitudes, assuming
a color of $J-K\sim2$ which is characteristic of an old passively evolving
system at $1<z<2$. The Y-axes show the $R-K$ color and the redshift (the
$R-K$ color being a function of $z$ for $1<z<2$). The continuous and dashed
diagonal lines indicate the location of $R=26$ and $R=28$ objects 
respectively. The vertical continuous line indicate the present limit
of seeing--limited near-IR {\it continuum} spectroscopy with VLT+ISAAC.
The vertical dashed line indicates a conservative limit of AO--aided 
near-IR continuum spectroscopy expected for the VLT. 
Different observing regimes can be envisaged. Continuum optical 
spectroscopy can be realistically performed down to $R\sim26$. Near-IR 
spectroscopy without AO can be done only for relatively bright targets 
($K\sim19$, $J\sim21$) and a $\geq1$ magnitude gain is expected with the 
aid of AO. For targets fainter than $K\sim20$, only imaging can be performed.  
For targets with, say, $R<28$, optical {\it and} near-IR
imaging can be done with reasonable integration times and photometric
accuracy, but for galaxies with $R>28$ optical imaging becomes very 
difficult and most of the photometric information comes from near-IR 
observations only.
}
\label{eps1}
\end{figure}

\item {\it AO--aided near-IR spectrograph.} The scientific
aim of such an instrument would be to spectroscopically confirm
the nature of the passive elliptical candidates and to measure
their redshifts. A simultaneous coverage of the near-IR spectral
range ($ZJHK$) would be essential to reduce the observation time
and to overcome the problem of matching and inter-calibrating independent
$Z,J,H,K$ spectra (e.g. ISAAC). A high throughput in the $Z$ and $J$
bands
would be also crucial to work efficiently in the spectral 
range  where the 4000~\AA~ break falls for $1.5<z<2.5$. A MOS capability
would be obviously important to increase the multiplex. Such a 
spectrograph should also have a 
low spectral resolution mode (e.g. R$\sim$200-500) in order to derive, 
together with optical photometry, accurate continuum spectrophotometry 
for the faintest targets in order to estimate their ``spectrophotometric'' 
redshifts whenever it is impossible to measure their spectroscopic 
redshifts (e.g. Cimatti et al. 1999; Soifer et al. 1999). A moderately 
high spectral resolution would be important to estimate 
the masses of the brightest targets through the velocity dispersion
of the absorption lines. It should be noted here that for faint galaxy 
spectroscopy (where the slit width cannot be much narrower than the 
size of the galaxy), low-order AO corrections are sufficient to provide 
a significant improvement of the S/N ratio thanks to reduction of 
the sky background compared with seeing-limited spectroscopy made with 
typical slit widths of 1$^{\prime\prime}$. 

\item {\it Near-IR wide-field imager} (e.g. 20$\times$20 arcmin$^2$). 
A VLT near-IR WFI with optimal seeing sampling (e.g. 0.15$^{\prime
\prime}$/pixel) would be crucial to exploit the image quality of 
the VLT and to push the photometry to the limits of a 8m-class 
telescope. Such an instrument would allow to perform ultradeep
surveys and to select and to study high-$z$ ellipticals beyond
the spectroscopic limits. In this respect, it would be essential
to have such an imager equipped with a set of medium-band filters
in order to derive spectral energy distributions and 
$z_{phot}$ with a high level of accuracy. Such an instrument
would play a key role in statistical studies such as the evolution
of the luminosity function and of the clustering of ellipticals. 
 
\end{itemize}

\section{Summary}

A significant population of $z>1$ passively evolving elliptical 
candidates was unveiled thanks to recent wide-field imaging surveys.
The confirmation of their nature and the measurement
of their redshifts are very challenging even with 8-10m class
telescopes and the results obtained so far are limited to the
brightest objects ($K<19$). Much observational work can still be 
done with the 1st generation VLT instruments such as the red-upgraded 
FORS2 (optical imaging and spectroscopy), CONICA and SINFONI (AO-aided 
near-IR imaging and spectroscopy).
An AO- or MCAO-aided near-IR imager-spectrograph (with MOS
capability) and a 
near-IR WFI equipped with medium-band filters are the most desirable 
2nd generation VLT instruments expected to play a crucial role
in the understanding of the formation and evolution of the
massive ellipticals.

%

\end{document}